\documentclass{aa}
\usepackage[varg]{txfonts}
\usepackage{hyperref}

\def\apjl{ApJ}
\def\aap{A\&A}

\def\nCIDobs{3\,273}
\def\nCID{3\,212}
\def\ConfCID{40}

\begin{document}
\title{Robust detection of CID double stars in SDSS}

\author{D.~Pourbaix\inst{1}\fnmsep\thanks{Senior Research Associate, F.R.S.-FNRS, Belgium}
  \and G.R.~Knapp\inst{2}
  \and J.E.~Gunn\inst{2}
  \and R.H.~Lupton\inst{2}
  \and \v{Z}.~Ivezi\'{c}\inst{3}
  \and C.~Siopis\inst{1}
  \and M.~Rigaux\inst{4}
\and A.~Rubbens\inst{5}}
\institute{Institut d'Astronomie et d'Astrophysique, Universit\'e Libre de Bruxelles (ULB), 1050 Brussels, Belgium  \email{pourbaix@astro.ulb.ac.be}
  \and
  Department of Astrophysical Sciences, Princeton University, Princeton, NJ 08544-1001, USA
  \and
  Department of Astronomy, University of Washington, Seattle, WA 98195-1580, USA
  \and
  D\'epartement de Physique, Universit\'e Libre de Bruxelles (ULB), 1050 Brussels, Belgium
  \and
  Coll\`ege Saint-Hubert, 1170 Brussels, Belgium
}
\date{Received 11 April 2016 / Accepted 11 May 2016}
\AANum{AA/2016/28688}
\abstract{}
{The Sloan Digital Sky Survey (SDSS) offers a unique possibility of not only detecting Colour Induced Displacement (CID) double stars but also confirming these detections.}
{Successive cuts are applied to the SDSS DR12 database in order to reduce the size of the sample to be considered.  The resulting dataset is then screened with a criterion based on the distance and orientation of the photocentres in different photometric bands.}
{About 3\,200 distinct objects are classified as CID double stars, 40 of which are confirmed with at least a second detection.  A consistency check further validates these detections.}
{}


\maketitle

\section{Introduction}\label{sec:intro}
The Sloan Digital Sky Survey \citep[SDSS;][ and references therein]{York-2000:a,Alam-2015:a} has revolutionised stellar astronomy since the late 90's by providing homogeneous and deep ($r < 22.5$) photometry in five passbands \citep[$u$, $g$, $r$, $i$, and $z$;][]{Fukugita-1996:a,Gunn-1998:a,Hogg-2001:a,Smith-2002:b,Doi-2010:a} accurate to 1-2\% \citep{Padmanabhan-2008:a}.  The sky coverage, 14\,555 deg$^2$ in the Northern Galactic Cap, results in photometric measurements for over 260 million stars and 208 million galaxies.  Astrometric positions are accurate to better than 0.1 arcsec per coordinate (rms) for point sources with $r<20.5^m$ \citep{Pier-2003:a}, and the morphological information from the images allows robust star-galaxy separation to $r \sim$ 21.5$^m$ \citep{Lupton-2003:a}.  The successive SDSS data releases \citep[e.g.,][]{Abazajian-2003:a,Abazajian-2004:a,Abazajian-2005:a,Adelman-McCarthy-2007:a} have provided the position of an increasing number of stars in the five photometric bands.  Up to SDSS DR7 \citep{Abazajian-2009:a}, for any given object, the positions at only one epoch were available.  Since SDSS DR8 \citep{Aihara-2011:a}, repeated observations have been reported.  

SDSS has been extensively used for extragalactic investigations, however other groups have taken advantage of it for stellar astrophysical purpose, especially binaries \citep{Silvestri-2007:a,Clark-2012:a}.  Thanks to the multiple photometric bands, colour-colour outlier detection was the first method adopted to filter the double stars out \citep{Raymond-2003:a,Smolcic-2004:a, Augusteijn-2008:a,Liu-2012:a}.  The photometric and spectroscopic capabilities of SDSS were also combined to detect photometrically well behaved objects \citep{Szkody-2002:a,Szkody-2003:a,Szkody-2003:b,Szkody-2004:a,Szkody-2005:a}, including post-common envelope binaries \citep{Schreiber-2010:a,Nebot-2011:a,Rebassa-2012:a}.  Spectroscopic binaries have also been detected thanks to the variability of their radial velocity \citep{Pourbaix-2005:a,Morganson-2015:a}.

SDSS pairs composed of a white dwarf and a main sequence (typically M) star have been quite intensively investigated \citep[see][and references therein]{2012MNRAS.423..320R} over the past 10 years.  These systems offer the combination of two stars at very distinct stages of their evolution but with a similar brightness, and distinctive colours.  Their value for our understanding of stellar evolution is therefore what also makes them rather easy to detect through unusual colours.

The detection of unresolved double stars through the wavelength dependence of the position of the photocentre (Colour Induced Displacement double stars, CID) was suggested by \citet{Christy-1983:a} and \citet{Sorokin-1985:a} and successfully applied to the SDSS DR2 and DR5 observations \citep{Pourbaix-2004:a,Pourbaix-2008:a}.  The same technique has lately been applied to the USNO-B1 dataset \citep{Jayson-2016:a}.  However, in all these investigations, the positions are measured at one epoch only, thus preventing any confirmation of what could simply be a false detection.  Although the detection of CID double stars requires 2+ photometric filters, it only relies upon the position measured through these filters, not the colour itself.  This method can therefore be applied to probe the whole stellar locus for binaries, with the exception of twins.  The more distinct the colour of the components, the farther apart their photocentres, so white dwarf + M dwarf are, again, among the privileged systems.  However, assuming the astrometry through the two filters can be tied up, there is no constraint on the type of the components as the detection capability is simply limited by the astrometric precision.

Taking advantage of the availability of repeated observations since DR8, we here present the multiple detection of CID double stars based on public SDSS data only.  The initial selection of the sample is described in Sect.~\ref{sec:observations}.  The CID criteria to be fulfilled are described in Sect.~\ref{sec:CID1results}.  The nature of the components is considered in Sect.~\ref{sec:CIDnature}. Finally, the time variability of the CID feature is analysed in Sect.~\ref{sec:CIDtime}.

\section{Observational data}\label{sec:observations}
All the data to be analysed come from one single table, PhotoObjAll, which contains more than one billion rows.  The number of rows has remained unchanged for the past five data releases but some bug fixes might have occurred so, from now on, DR12 data \citep{Alam-2015:a} are going to be assumed.  A significant improvement introduced in DR9 from the viewpoint of this investigation is related to the differential chromatic correction (DCR).  \citet{Pier-2003:a} describe how DCR is calibrated and corrected but the DR9 astrometry page\footnote{https://www.sdss3.org/dr9/algorithms/astrometry.php} mentions that DCR has been fully accounted for from DR9 on only.  The results of \citet{Pourbaix-2008:a} were therefore still affected by some colour terms unaccounted for.

Successive selections are required to clean up the sample initially composed of PhotoObjAll.  Only the observations of objects belonging to the Star view are considered.  The same filtering as introduced by \citet{Pourbaix-2004:a} was applied: an observation was removed if any of the flags {\em saturated}, {\em bright}, {\em edge}, or {\em nodeblend} was set or if the precision on $u$, $g$, $r$, $i$, or $z$ was larger than 0.1, 0.05, 0.1, 0.05, and 0.05 respectively.

There are 25\,797\,735 stars that fulfil these criteria.  For them, the standard deviation of the offsets with respect to the $r$ position in $z$ for both $\alpha\cos\delta$ and $\delta$ are 21 mas.  For the $u$ band, the standard deviation in $\alpha\cos\delta$ and $\delta$ are 42 mas and 44 mas respectively.  These four values are based on the central 99\% of the offsets.  Despite this 1\% clipping, the scatter of the offsets in $u$ is 30\% larger than derived by \citet{Pier-2003:a}.  Both the right ascensions (together) and declinations (together) are correlated at the level 0.16 whereas the correlation between any right ascension and any declination is always 1 order of magnitude smaller.

\section{CID filtering}\label{sec:CID1results}

\begin{figure}[htb]
\begin{center}
  \resizebox{\hsize}{!}{\includegraphics{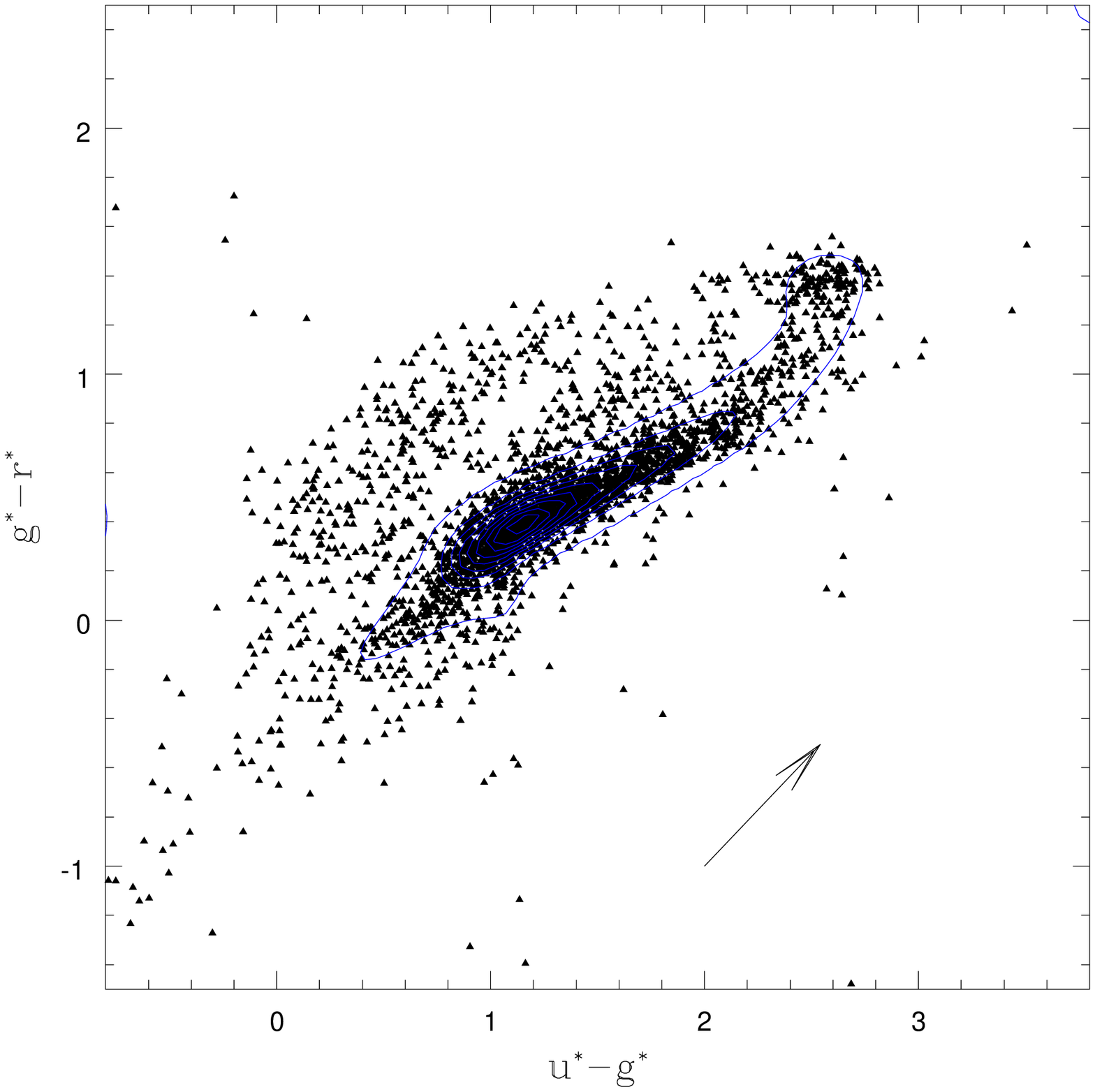}}\\
\resizebox{\hsize}{!}{\includegraphics{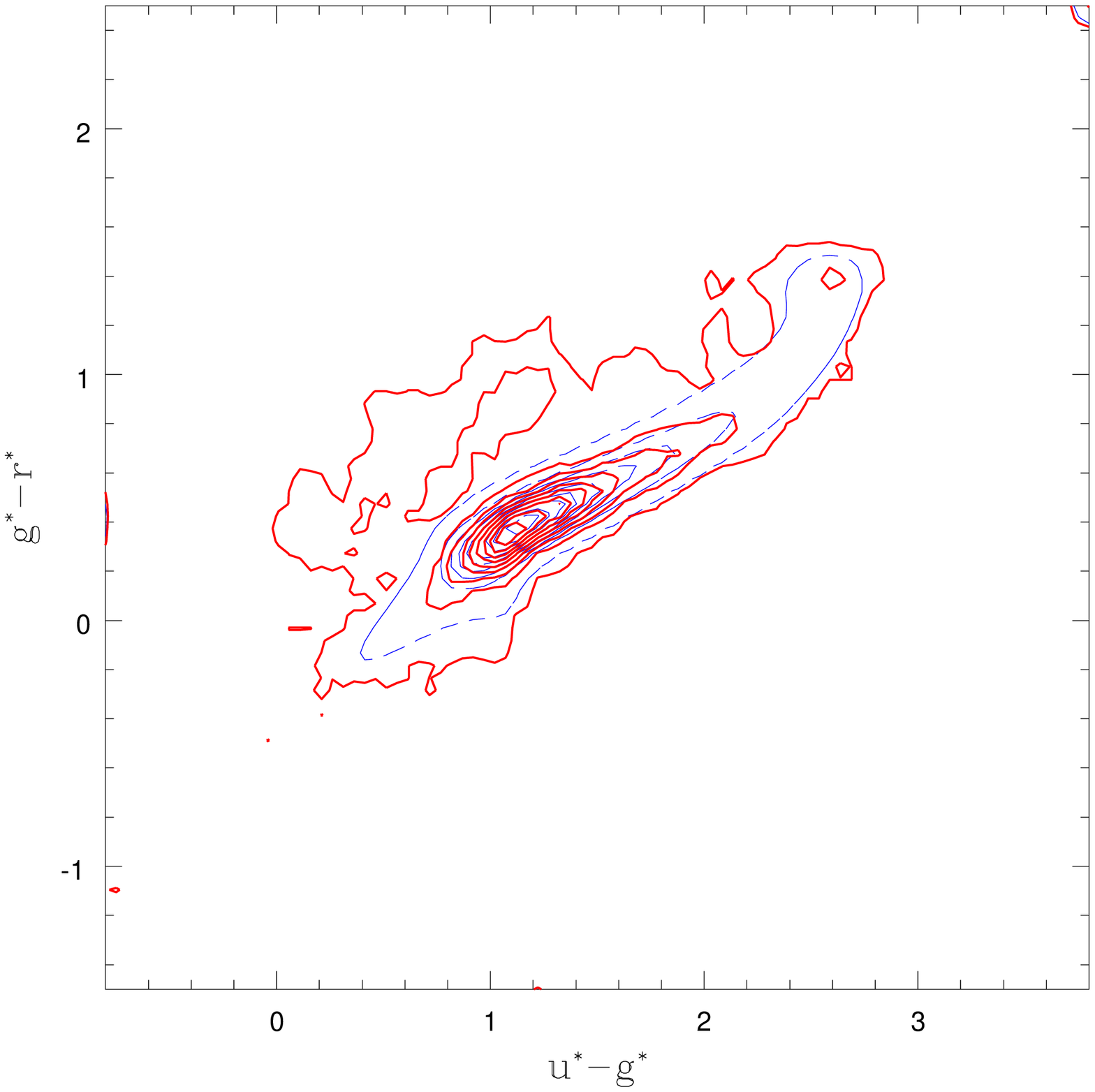}}
\end{center}
\caption[]{\label{fig:colcolCID}(Top panel) Position of the CID candidates in the dereddened colour-colour diagram together with the stellar locus based on 26 million Star colours (i.e. 10\% of the stellar content of SDSS).  The arrow denotes the extinction \citep{2011ApJ...737..103S}.  (Bottom panel) Contour lines for the CID candidates (thick) and 10\% of the stellar content of SDSS (thin dashed).}
\end{figure}

The idea behind Colour Induced Displacement double stars is that the wavelength-dependent photocentres should be aligned along the two stars rather than being randomly distributed around some median position \citep{Pourbaix-2004:a}.  Even though the positions in two photometric bands are enough to notice a displacement \citep{Jayson-2016:a}, three positions are necessary to assess the alignment.  Among the five photometric bands from SDSS, $u$ and $z$ are the furthest apart in terms of wavelength, whereas $r$ is more central.

The results of the previous section used in a simulation reveal that about 1 single star out of 600\,000 could accidentally have its $u$ and $z$ positions more than 0.4\arcsec\ apart and aligned to more than 177.5\degr. (actually, cosine lower than -0.999).  In a sample of 10 million observations, we thus anticipate 16 such false detections.  However, the probability of the same single star to be observed twice with the same features on its positions is about $3\,10^{-12}$. 

There are \nCIDobs\ stellar observations which match the criteria on separation ($>=0.4$\arcsec), orientation ($>=177.5$\degr) and photometric quality.  They correspond to \nCID\ distinct objects (Table~\ref{tab:CIDcandidates}) whose accepted CID observations are listed in Table~\ref{tab:CIDobservations}.  The location of these points in a dereddened colour-colour diagram is plotted in Fig.~\ref{fig:colcolCID}.  For the sake of comparison, the stellar locus based on 10\% of the Star entries, with the same photometric properties (ranges and precisions) as the CID candidates is plotted as well.  Even though the photometric distribution of the CID candidates essentially overlap with the locus of the regular stars, it also leaks on the upper left region of the latter (Fig.~\ref{fig:colcolCID}).  We shall come back to this point in Sect.~\ref{sec:CIDnature}.

\setlength{\tabcolsep}{1pt}
\begin{table}[htb]
  \caption[]{\label{tab:CIDcandidates}Identified CID with their thingId, position, and SDSS dereddened magnitude for the five photometric bands.  $N$ is the number of observations flagged as CID.  The whole table is available in electronic version only.}
  \begin{tabular}{lrrcccccc}
    \hline\hline
    thingId & RA(\degr) & Dec (\degr) & $N$ & $u^*$ & $g^*$ & $r^*$ & $i^*$ & $z^*$ \\
    \hline
15431000  & 212.8479 & -13.1359 & 2 & 19.60 & 18.32 & 17.58 & 16.68 & 16.14\\
\hline
\end{tabular}
\end{table}

\begin{table*}[htb]
  \caption[]{\label{tab:CIDobservations}Accepted CID observations for the candidates listed in Table~\ref{tab:CIDcandidates} with their thingId, objectId, date, and offsets in right ascension ($\alpha^*=\alpha\cos\delta$), and declination in the $u$ and $z$ bands with respect to the position in the $r$ band.}
  \begin{tabular}{lllcccc}
    \hline\hline
    thingId & objectid & MJD & $\Delta\alpha^*_u$ & $\Delta\delta_u$ & $\Delta\alpha^*_z$ & $\Delta\delta_z$ \\
            &          &     & (\arcsec) & (\arcsec) & (\arcsec) & (\arcsec)\\
    \hline
   523314 & 1237668627303039488 & 53527.23 & -0.0081 & +0.0111 & +0.2453 & -0.3173 \\
\hline
\end{tabular}
\end{table*}

In terms of distribution over the sky, there is a small excess of CID detections close to the equator with respect to the parent population.  That is however consistent with a criterion partly based on $\Delta\alpha\cos\delta$ where $\Delta\alpha$ stands for the difference of offsets in right ascension in the $u$ and $z$ bands.  For an object close to the equator, that quantity and therefore the separation between the $u$ and $z$ photocentres are more likely to exceed any adopted threshold.

\begin{figure}[ht]
\begin{center}
\resizebox{\hsize}{!}{\includegraphics{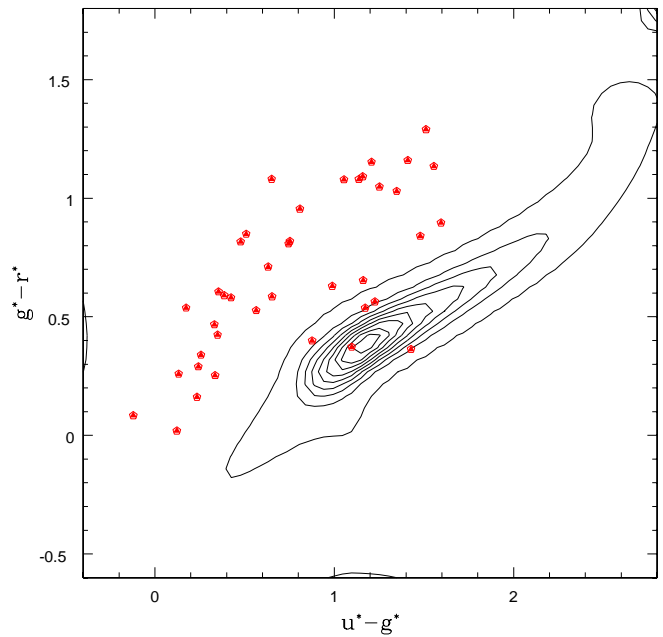}}
\end{center}
\caption[]{\label{fig:colcolconfCID}Position of the confirmed CID candidates in the dereddened colour-colour diagram together with the stellar locus.}
\end{figure}

So far, the adopted methodology is the same as in \citet{Pourbaix-2004:a} and the change in the number of candidates is only caused by the substantial increase of the number of stars observed.  Since DR8, some stars have been observed on several occasions, making possible the search for CID candidates among them.  It is worth noting that a detection does not necessarily mean a photometric observation (valid or not).  The number of observations stored in PhotoObjAll thus often turns out to be lower than publicised by the field {\em nDetect}.  For instance, if one considers the stars detected 5+ times, they account for nearly 76 million detections but PhotoObjAll only contains 66 million observations for them.  Further imposing that a star was detected more than once, we are left with \ConfCID\ objects with 2+ CID-like observations (i.e. with a much lower risk of being false detections).  The position of these confirmed CID candidates in the colour-colour diagram together with the stellar locus is plotted in Fig.~\ref{fig:colcolconfCID}.  Their location on the sky is listed in Table~\ref{tab:confCID}.

\setlength{\tabcolsep}{3pt}
\begin{table}[htb]
  \caption[]{\label{tab:confCID}Confirmed CID candidates where thingId denotes the unique SDSS identifier, RA and Dec are the right ascension and declina\-tion (2000.0).  N is the number of observations flagged as CID.  Spec is the optical spectral classification listed by the DR12 Science Archive Server, * indicates the presence of some prominent Balmer lines.  The link points directly to the image of the object on DR12 SkyServer.  The URL of the image is http://skyserver.sdss.org/dr12/en/tools/chart/navi.aspx?ra=RA\&dec=Dec where RA and Dec are the values listed in the second and third columns.  The live links are available in the on-line version of this paper.}
  \begin{tabular}{lrrlll}
    \hline\hline
    thingId & RA(\degr) & Dec (\degr) & N & Spec & Image \\
    \hline
15431000  & 212.847899 & -13.135926 &2&      & \href{http://skyserver.sdss.org/dr12/en/tools/chart/navi.aspx?ra=212.847899646762&dec=-13.1359263119831}{SkyServer}\\
22550603  & 117.733729 & -10.479184 &2&      & \href{http://skyserver.sdss.org/dr12/en/tools/chart/navi.aspx?ra=117.733729777588&dec=-10.4791846419856}{SkyServer}\\
45706433  & 324.637702 &  -5.054417 &2&      & \href{http://skyserver.sdss.org/dr12/en/tools/chart/navi.aspx?ra=324.637702303223&dec=-5.05441696472427}{SkyServer}\\
47860353  & 324.727576 &  -4.391773 &2&      & \href{http://skyserver.sdss.org/dr12/en/tools/chart/navi.aspx?ra=324.727576075877&dec=-4.39177349560784}{SkyServer}\\
68189740  &   5.546124 &  -1.128588 &3&   M4 & \href{http://skyserver.sdss.org/dr12/en/tools/chart/navi.aspx?ra=5.5461429&dec=-1.1286179}{SkyServer}\\
69946683  & 320.716211 &  -1.097230 &2&      & \href{http://skyserver.sdss.org/dr12/en/tools/chart/navi.aspx?ra=320.716211667961&dec=-1.09723079708579}{SkyServer}\\
74734108  &  94.881818 &  -0.996007 &2&      & \\
83883365  & 323.069439 &  -0.426253 &3&      & \href{http://skyserver.sdss.org/dr12/en/tools/chart/navi.aspx?ra=323.069439511884&dec=-0.426253213274254}{SkyServer}\\
84696684  &  57.788778 &  -0.480402 &2&      & \\
85303451  & 351.483366 &  -0.499383 &2&      & \\
93807459  &  70.608754 &  -0.206033 &3&      & \href{http://skyserver.sdss.org/dr12/en/tools/chart/navi.aspx?ra=70.6087542129561&dec=-0.206033067889766}{SkyServer}\\
94010862  &  47.522422 &  -0.050319 &2&      & \\
94025347  &  94.805658 &  -0.158900 &2&      & \href{http://skyserver.sdss.org/dr12/en/tools/chart/navi.aspx?ra=94.8056580100082&dec=-0.158900454839095}{SkyServer}\\
96981424  &  17.849609 &   0.159764 &5&   M2*& \href{http://skyserver.sdss.org/dr12/en/tools/chart/navi.aspx?ra=17.849581&dec=0.1598125}{SkyServer}\\
103499550 &  74.703028 &   0.215468 &2&      & \\
106358786 &  15.923339 &   0.525698 &3&   M2*& \\
107874362 & 346.798161 &   0.477047 &2&      & \href{http://skyserver.sdss.org/dr12/en/tools/chart/navi.aspx?ra=346.798161533339&dec=0.477047597165135}{SkyServer}\\
115775052 &  13.577270 &   0.962821 &4&   M4*& \href{http://skyserver.sdss.org/dr12/en/tools/chart/navi.aspx?ra=13.5772700334341&dec=0.962821822084125}{SkyServer}\\
116724303 &  81.454729 &   1.005443 &4&      & \href{http://skyserver.sdss.org/dr12/en/tools/chart/navi.aspx?ra=81.4547299331792&dec=1.00544392941365}{SkyServer}\\
116891499 & 358.869452 &   1.006782 &2&   M1 & \\
116995966 & 348.090035 &   1.024174 &5&   M3*& \href{http://skyserver.sdss.org/dr12/en/tools/chart/navi.aspx?ra=348.09&dec=1.024175}{SkyServer}\\
119480169 & 324.594285 &   1.062909 &2&      & \\
121325158 &  15.037479 &   1.142725 &4&   M1*& \\
121351015 &  21.482147 &   1.075074 &4&      & \\
121561453 &  55.658241 &   1.149527 &3&   M3*& \href{http://skyserver.sdss.org/dr12/en/tools/chart/navi.aspx?ra=55.6582412192943&dec=1.14952751343703}{SkyServer}\\
122233016 &   9.736654 &   1.114130 &3&   M4 & \\
132088747 & 261.203461 &   2.253603 &2&      & \\
184456091 &  27.072581 &   8.015645 &2&      & \href{http://skyserver.sdss.org/dr12/en/tools/chart/navi.aspx?ra=27.0725815968507&dec=8.01564557255167}{SkyServer}\\
186374149 & 203.360705 &   8.254411 &2&      & \href{http://skyserver.sdss.org/dr12/en/tools/chart/navi.aspx?ra=203.360705933828&dec=8.25441108363379}{SkyServer}\\
200646968 & 126.409171 &   9.857400 &2&      & \href{http://skyserver.sdss.org/dr12/en/tools/chart/navi.aspx?ra=126.409171258216&dec=9.85740076447783}{SkyServer}\\
220442434 & 125.558020 &  12.243735 &2&   M3*& \href{http://skyserver.sdss.org/dr12/en/tools/chart/navi.aspx?ra=125.55803&dec=12.243746}{SkyServer}\\
259525347 & 251.906928 &  16.610724 &2&      & \href{http://skyserver.sdss.org/dr12/en/tools/chart/navi.aspx?ra=251.906928488818&dec=16.6107243705808}{SkyServer}\\
274459969 & 236.094783 &  18.549545 &2&      & \href{http://skyserver.sdss.org/dr12/en/tools/chart/navi.aspx?ra=236.094783053253&dec=18.549545482691}{SkyServer}\\
337204518 & 120.918335 &  25.924343 &2&      & \\
395512014 & 341.514497 &  33.791220 &3&      & \href{http://skyserver.sdss.org/dr12/en/tools/chart/navi.aspx?ra=341.514497230845&dec=33.7912204905931}{SkyServer}\\
405351640 & 121.363021 &  35.285873 &2&      & \\
429504180 & 253.606452 &  39.296835 &2&      & \href{http://skyserver.sdss.org/dr12/en/tools/chart/navi.aspx?ra=253.606452061869&dec=39.2968356882463}{SkyServer}\\
451341367 & 122.130724 &  43.010222 &2& M4.5III*& \\
486944679 & 247.731754 &  50.049775 &2&      & \href{http://skyserver.sdss.org/dr12/en/tools/chart/navi.aspx?ra=247.731754914187&dec=50.0497754254427}{SkyServer}\\
490371322 & 133.634605 &  50.857572 &2& M4.5III*& \href{http://skyserver.sdss.org/dr12/en/tools/chart/navi.aspx?ra=133.634605474341&dec=50.8575727133433}{SkyServer}\\
    \hline
  \end{tabular}
\end{table}

\section{Tentative nature of the components}\label{sec:CIDnature}

As already stated in Sect.~\ref{sec:CID1results}, the ($u^*-g^*$,$g^*-r^*$) locus of the CID overlaps with the single star one with a distinctive leakage towards the upper left corner of the latter.   The contour line shows that the flooding is not continuous but instead represents a bridge between the M and white dwarfs \citep{Smolcic-2004:a}.  The three CID depicted in Fig.~\ref{fig:monotonicEvol} belong to this category.

A second leakage of the single star locus on its lower left end results from an overcorrection of the galactic extinction \citep{1998ApJ...500..525S} for objects close to the galactic plane.  Regardless of how unrealistic some of these dereddened colours are, they cannot be responsible for the CID status.  Indeed, the criteria to be considered as a CID are purely astrometric.  Even though some spurious astrometric results can sometime come from a wrong chromatic correction \citep{Pourbaix-2003:b}, any chromatic correction is based on the observed colours, not the dereddened ones.  So, even if some dereddened colours might be wrong, the positions should nevertheless always be accurate.

With respect to the single star stellar locus, the CID candidates exhibit an excess of M-dwarf-like objects centred in (2.6, 1.4) on the ($u^*-g^*$,$g^*-r^*$) density map (Fig.~\ref{fig:colcolCID}).  According to \citet{2008AJ....135..785W}, their $riz$ colours correspond to spectral types ranging from M0 to M4.

Among the \ConfCID\ confirmed CID, only twelve got their spectral type directly determined through spectroscopy \citep{2002AJ....123..485S,2012AJ....144..144B}.  All are M-type: ten dwarfs and two giants.  In seven cases, some prominent Balmer lines are detected in absorption, hints of the presence of a white dwarf in the same spectroscopic field of view.  Over the years and the data releases, several groups have published some lists of white dwarfs, M-dwarfs, and binaries with a white dwarf \citep{2009A&A...496..191H,2010MNRAS.402..620R,2011AJ....141...97W,2013ApJS..204....5K}.  In total, ten of our candidates belong to at least one of these lists.  The remaining 30 objects (including the two with an M giant component) are still completely absent from any published investigation (according to Simbad).  In particular, none of the five objects which belong to the stellar locus in Fig.~\ref{fig:colcolconfCID} got its spectral type determined in the framework of SDSS, nor by any other investigation.

\begin{figure}[ht]
\begin{center}
\resizebox{\hsize}{!}{\includegraphics{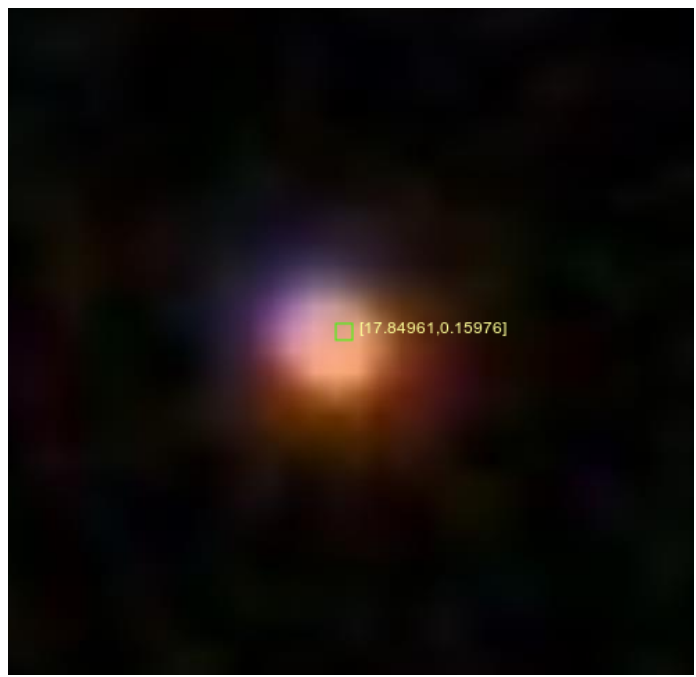}}
\end{center}
\caption[]{\label{fig:Image}The image of 96981424 (SDSS J011123.90+000935.1) exhibits a colour gradient whose orientation seems to be orthogonal to the CID.}
\end{figure}

  For twenty five objects for which an image is available, that image, though point-like, clearly shows two regions of distinct colours, thus leading to a distinct position for the photocentre in at least two photometric bands (e.g. thingId \#96981424, Fig.~\ref{fig:Image}).  All the individual images, at their highest resolution, are directly accessible through the link {\em SkyServer} in Table~\ref{tab:confCID}.

\section{CID over time}\label{sec:CIDtime}

Even when 2+ observations secure the CID status of an object, that feature is far from being present in all its observations.  One can therefore wonder how robust these detections are, even when confirmed at least once.  So far, in this paper, only a tightened version of the criterion from the 2004 investigation has been used.  Carrying out a consistency check among the successive CID-like observations would already be a substantial improvement.  

\begin{figure}[ht]
\begin{center}
\resizebox{\hsize}{!}{\includegraphics{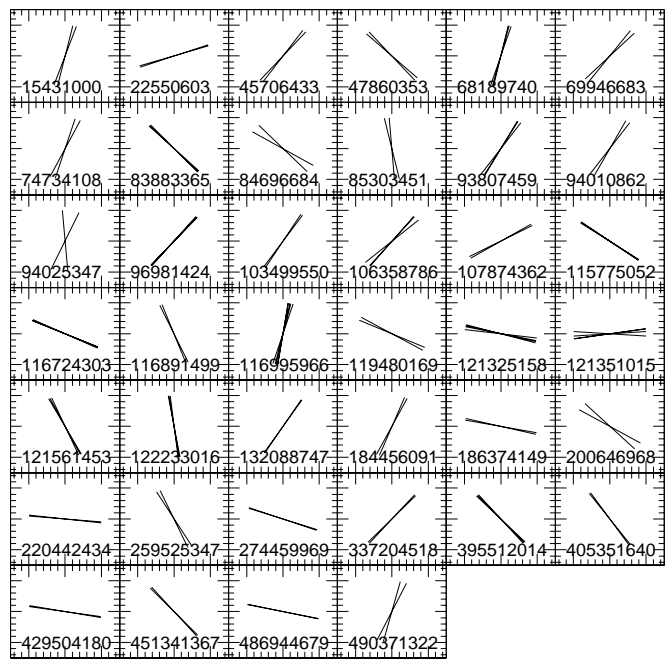}}
\end{center}
\caption[]{\label{fig:orientation}Orientation of the lines joining the $u$ and $z$ positions of the \ConfCID\ double stars with 2+ CID-like observations.  The numbers are the SDSS {\em thingid} field.}
\end{figure}

The line between the $u$ and $z$ positions essentially represent the line between the two stars.  Unless a CID-candidate double star turns to be a short period binary (the observations cover at most 4178 days), the relative orientation based on the $u$ and $z$ positions should be fairly stable over time.  For each of the \ConfCID\ confirmed CID double stars, the orientations of all the validating observations are plotted in Fig.~\ref{fig:orientation}.

\begin{figure}[ht]
\begin{center}
\resizebox{\hsize}{!}{\includegraphics{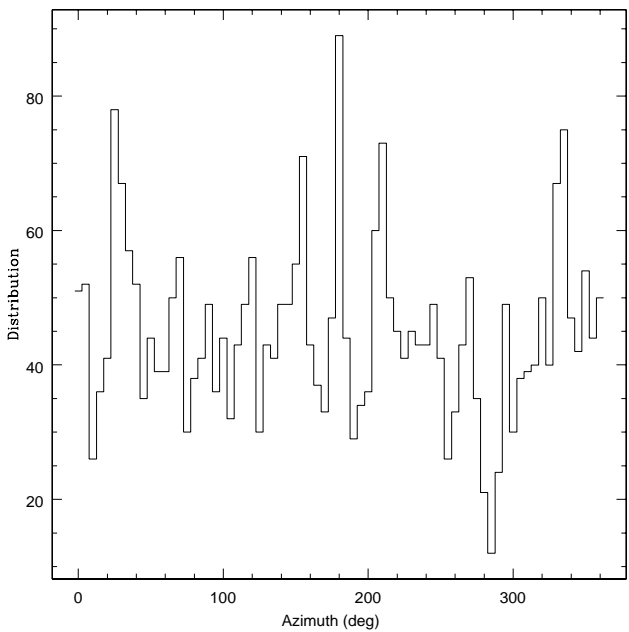}}
\end{center}
\caption[]{\label{fig:orientationNonConfirmed}Distribution of the orientations of the lines joining the $u$ and $z$ positions of all the CID-like observations.}
\end{figure}

A vast majority of the CID candidates are unfortunately not confirmed by a second observation but one can nevertheless assess their global behaviour.  For instance, the azimuth of all the CID candidates should be uniformly distributed over 0--360\degr.  A $\chi^2$-test performed on the binned azimuths (Fig.~\ref{fig:orientationNonConfirmed}) rejects the uniformity at the 99\% confidence level.  In order to assess whether this behaviour points towards any instrumental effect or not, a similar analysis was performed on the position angles of the 11\,057 members of the Double and Multiple Star/Component solutions of Hipparcos and Tycho Catalogues \citep{Hipparcos}.  At the same confidence level, the uniformity hypothesis of the azimuths of the Hipparcos resolved pairs is not rejected.   So, why is it rejected with the SDSS CID?   A Monte-Carlo simulation directly rules out any explanation based on the size of the sample.  

Regardless of the coordinate and the filter, the four offsets of the CID observations are symmetrically distributed around 0.  However, whereas the distribution of the difference of offset in declination is bi-modal, the $\alpha*$ version, though similar, exhibits a third peak right on 0.  For all the objects in that third peak, the corresponding azimuth is either 0 or 180\degr\, regardless of the offsets in declination, thus contributing to breaking the uniformity hypothesis.  In terms of offset in declination, there is a tiny departure from symmetry right after 0, thus causing a lack of azimuths around 280 degrees.  The objects identified in these bins do not share any common location on the sky.

\begin{figure*}[ht]
\begin{center}
\resizebox{0.32\hsize}{!}{\includegraphics{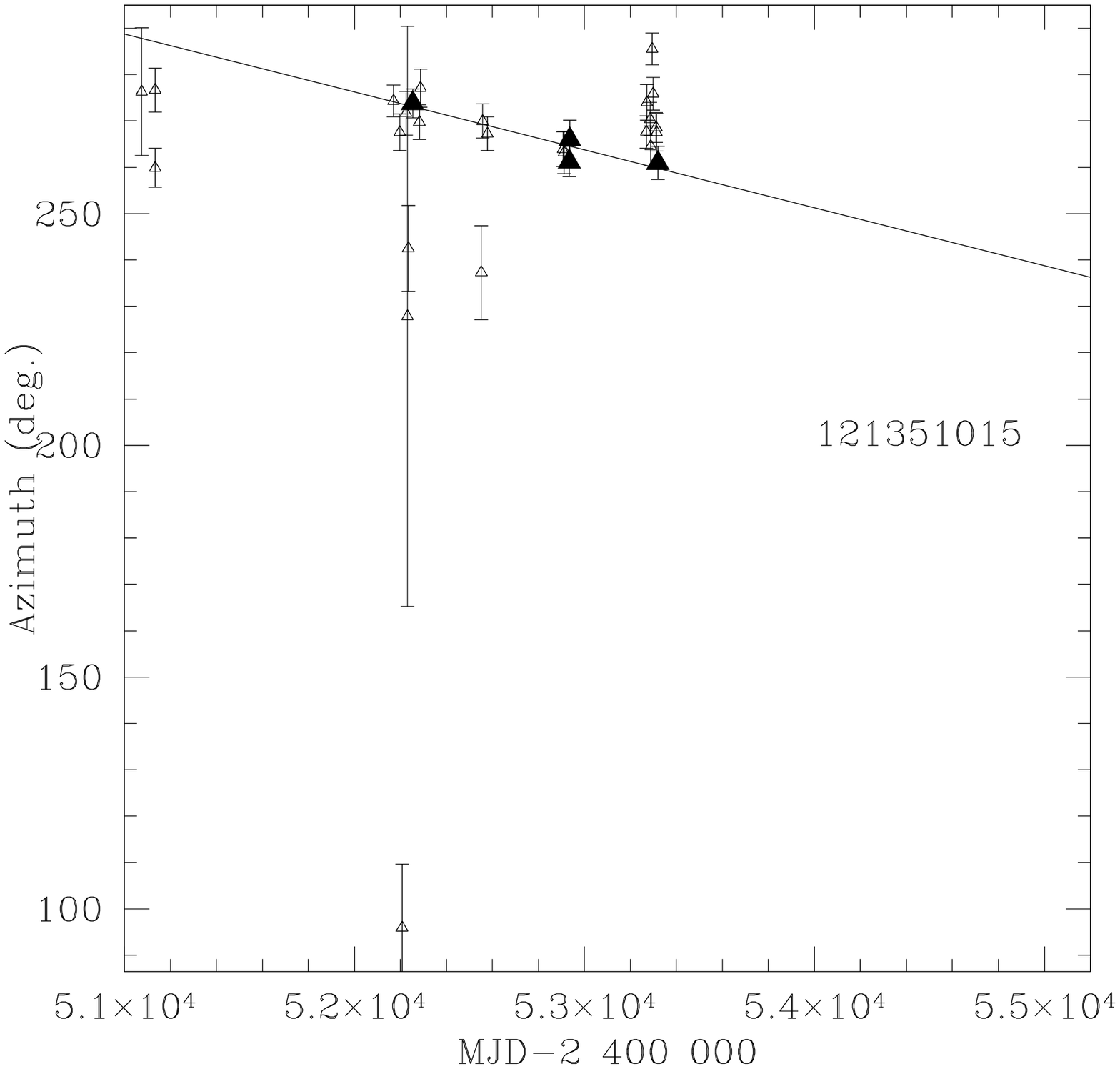}}
\resizebox{0.32\hsize}{!}{\includegraphics{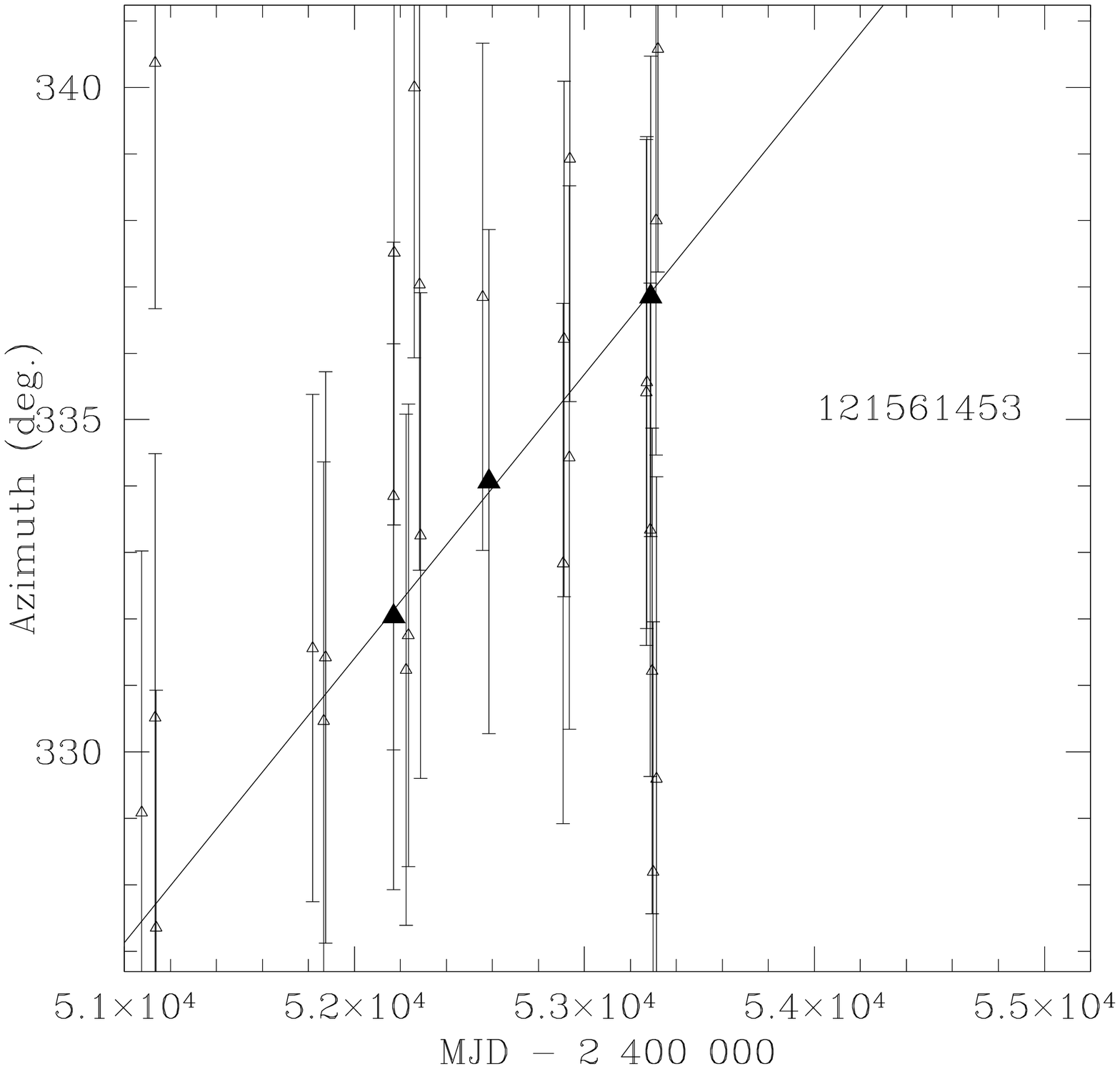}}
\resizebox{0.32\hsize}{!}{\includegraphics{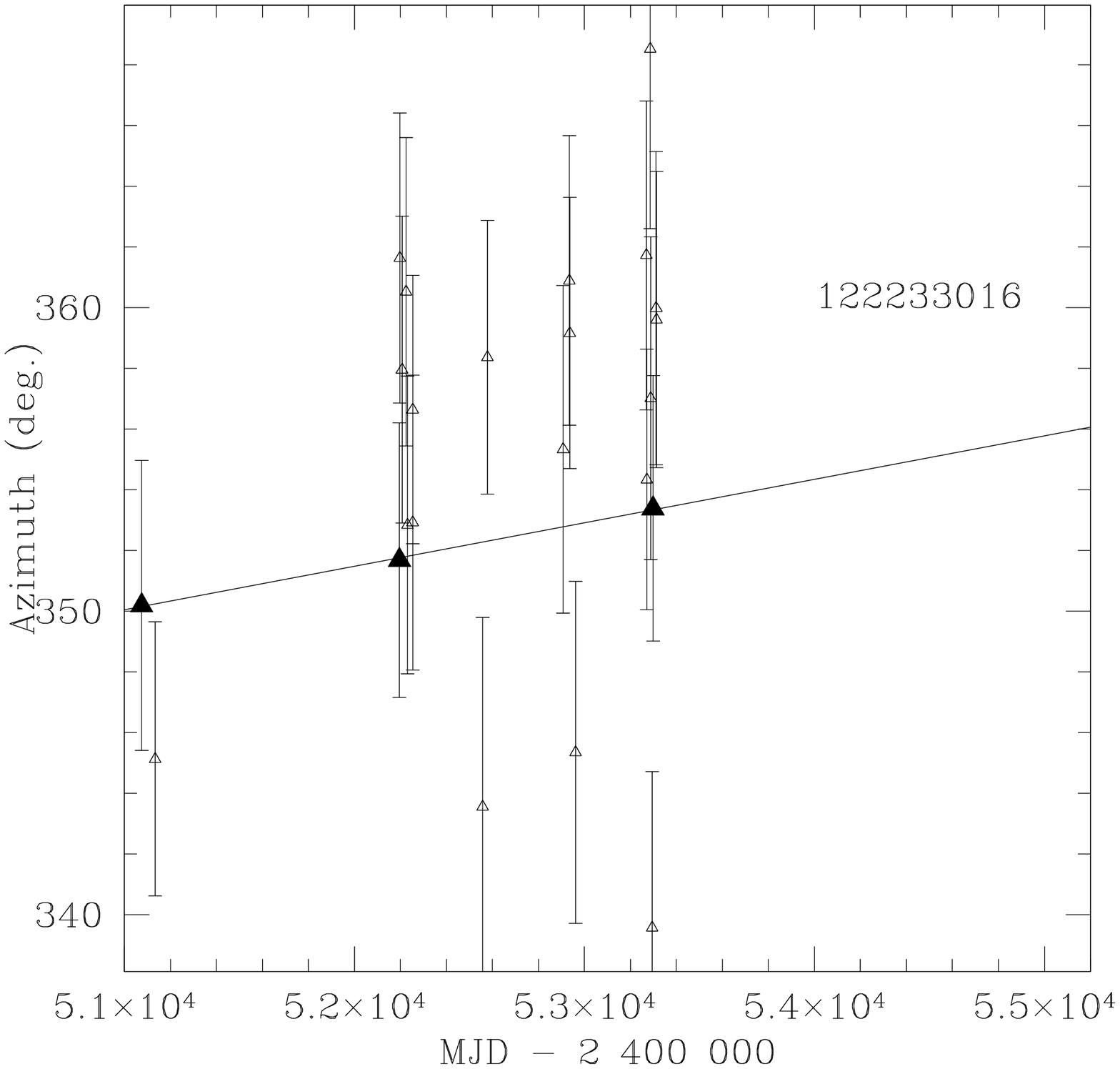}}
\end{center}
\caption[]{\label{fig:monotonicEvol}Evolution of the azimuth over time.  Open triangles denote observations without CID status.  Filled triangles are valid CID observations and only they were included in the fit.  The number is the unique thingId SDSS object number.}
\end{figure*}

Although Fig.~\ref{fig:orientation} offers a good assessment of the CID nature of most of the \ConfCID\ candidates, the successive observations do not provide any confirmation that any of these double stars is a binary.  However, among the thirteen CID with 3+ valid observations, there are three objects which also exhibit a monotonic evolution of the azimuth (Fig.~\ref{fig:monotonicEvol}).  Actually, the evolution is not just monotonic: the azimuth of the valid CID observations follows a linear function of the time corresponding to $-4.\pm1.3$, $1.56\pm0.092$, and $0.52\pm0.022$ deg\,yr$^{-1}$ respectively.

Inferring any orbital period from this azimuth gradient would be extremely speculative.  Object 12561453 (SDSS J034237.97+010858.2) was modelled as a DA white dwarf + M3 dwarf system \citep{Rebassa-2012:a}, $339\pm84$pc away from us.  Using their tentative masses and assuming an angular separation of 0.7\arcsec\ (which is a lower bound based on the $u$ and $z$ positions) yield a period of $3800\pm1500$yr to be compared with the $230\pm15$yr we obtain assuming that our rate remains constant.

Could this evolution of the azimuth be the sign of a genuine binary?  Generally speaking, no, unless the orbit is seen face on and the epochs are adequately distributed.  However, for the three cases, the change of the azimuth does not exceed a couple of degrees over 1\,000+ days.  For any orbital segment short enough, such a linear evolution of the azimuth would be very likely.  Regardless of how appealing this explanation sounds, an alternative would be two stars passing next to each other in the plane perpendicular to the line of sight.  Once again, such a motion would not yield a linear change of the azimuth in general except for the short path for which the linear approximation holds.  As already stated in the title of this paper, CID are double stars, not necessarily binaries.  One cannot exclude that they are made up of two stars accidentally on the same line of sight, although being far apart.  However, if that is true for this technique, that is also true for the objects detected through their peculiar colours.

\section{Conclusions}\label{sec:conclusions}

Whereas the possibility of detecting CID double stars with SDSS has been known for more than a decade, the absence of repeated observation (in particular of CID candidates) has prevented the validation of the technique, despite its easy setup.  We have shown here that some CID are confirmed by at least a second observation.  Furthermore, a consistency check confirms, with a criterion completely independent of the CID one, that the position angle of one component with respect to the other is stable over the time covered by the SDSS data (less than 4\,000 days).  In three cases, a linear evolution of the azimuth of the stars is noticed over at least three distinct epochs.  However, one cannot rule out the possibility that a high proper motion star is apparently passing by a distant star.

\begin{acknowledgements}
DP thanks Alain Jorissen for the stimulating discussions about the astrophysical aspects of these objects and A.~Thakar from the SDSS helpdesk for his support in bypassing some early CasJobs limitations.  We thank the referee, Andrei Tokovinin, for his valuable comments and suggestions.  Funding for SDSS-III has been provided by the Alfred P. Sloan Foundation, the Participating Institutions, the National Science Foundation, and the U.S. Department of Energy Office of Science. The SDSS-III web site is http://www.sdss3.org/.

SDSS-III is managed by the Astrophysical Research Consortium for the Participating Institutions of the SDSS-III Collaboration including the University of Arizona, the Brazilian Participation Group, Brookhaven National Laboratory, Carnegie Mellon University, University of Florida, the French Participation Group, the German Participation Group, Harvard University,
Instituto de Astrof\'isica de Canarias,
the Michigan State/Notre Dame/JINA Participation Group, Johns Hopkins University, Lawrence Berkeley National Laboratory,
Max-Planck-Institut f\"ur Astrophysik (MPA Garching), 
Max-Planck-Institut f\"ur Extraterrestrische Physik (MPE), 
New Mexico State University, New York University, Ohio State University, Pennsylvania State University, University of Portsmouth, Princeton University, the Spanish Participation Group, University of Tokyo, University of Utah, Vanderbilt University, University of Virginia, University of Washington, and Yale University.  This research has made use of "Aladin sky atlas" developed at CDS and of the SIMBAD database, operated at CDS, Strasbourg Observatory, France.
\end{acknowledgements}

\bibliographystyle{aa}
\bibliography{articles,books}

\end{document}